\documentclass[12pt]{iopart}
\usepackage{epsfig}

\begin{document}

\title
{Entanglement of topological phase factors}

\author{D. I. Tsomokos}
\address{Department of Computing, University of Bradford, Bradford BD7 1DP, UK}
\address{Department of Physics, Astrophysics and Mathematics,
University of Hertfordshire, College Lane, Hatfield, AL10 9AB, UK}

$\\$

\emph{New Journal of Physics} \textbf{7}, 50 (2005)

\begin{abstract}
The topological phase factor induced on interfering electrons by external
quantum electromagnetic fields has been studied. Two and three electron
interference experiments inside distant cavities are considered and the
influence of correlated photons on the phase factors is investigated. It is
shown that the classical or quantum correlations of the irradiating photons
are transferred to the topological phases. The effect is quantified in terms
of Weyl functions for the density operators of the photons and illustrated
with particular examples. The scheme employs the generalized phase factor as a
mechanism for information transfer from the photons to the electric charges.
In this sense, the scheme may be useful in the context of flying qubits
(corresponding to the photons) and stationary qubits (electrons), and the
conversion from one type to the other.
\end{abstract}

\maketitle

\section{Introduction}
The study of phase factors arising in quantum interference has been crucial
for the understanding of a wide range of physical phenomena
\cite{Geometric_phases}. The Aharonov-Bohm phase factor \cite{AB}, $\exp(\rmi
q\Phi)$, is acquired by a particle with charge $q$ in a looping trajectory
that encloses a classical magnetostatic flux $\Phi$. This is true even when
the particle moves in entirely field-free regions. The effect has been
investigated in relation to transport phenomena in solid state physics
\cite{SS} and electron coherence in mesoscopic devices \cite{WP}. The
reciprocal phase factor \cite{AC} and the dual counterparts
\cite{Wilkens_phase,Dowling} have also been studied and have recently found
applications in different contexts, such as topological quantum information
processing \cite{AB_gates}, the quantum Hall effect analogue with neutral
atoms \cite{Ericsson} and ultra-cold atom technology \cite{Pachos}.

The generalized phase factor, $ \exp(\rmi q\hat\phi) $, which is induced on a
charge $q$ by a nonclassical electromagnetic field with magnetic flux $\hat
\phi$ has also been studied in the literature \cite{Vourdas_AB}. In this case
the magnetic flux and the induced phase factor are quantum mechanical
operators. Consequently the important quantity in terms of interference
properties is the expectation value of the phase factor, $\langle \exp(\rmi
q\hat\phi) \rangle = \Tr [\rho \exp(\rmi q\hat\phi)]$, with respect to the
density matrix $\rho$ that describes the external electromagnetic field. This
phase factor is topological in the sense that it depends on the number of
times an electron winds around the enclosed magnetic flux and it is
independent of the electron velocity. The $\langle \exp(\rmi q\hat\phi)
\rangle $ is a complex quantity, in general, which is known as the Weyl (or
characteristic) function from quantum phase-space studies \cite{WW}.

Clearly the inherent fluctuations of the external quantum fields bring about
the problem of decoherence of the interfering electrons. Solutions have been
proposed in relation to this problem using various methods \cite{Ford_AB}.
Here it is assumed that under certain conditions the external photons do not
interact with the interfering charges. In particular, it is assumed that the
electromagnetic fields that are induced via Faraday's law by the circulating
electrons are negligible in comparison to the external fields, and so there is
no back reaction. The inherent noise of the external photons manifests itself
as a reduction of the absolute value of the phase factor, $|\langle \exp(\rmi
q\hat\phi) \rangle|$, which becomes slightly less than one \cite{Chong}.

Nonclassical electromagnetic fields in various quantum states \cite{Loudon},
such as squeezed and number states, have been generated at both optical and
microwave frequencies \cite{em_fields}. Quantum mechanically correlated
\cite{EN} photons have also been produced in the laboratory \cite{EN_photons}.
It is therefore reasonable to enquire whether we can use certain quantum
interference devices, which are sensitive to the external radiation, as
detectors of photon correlations. This has indeed been proposed recently using
different techniques \cite{Related_MY,MY}. In this paper we study
photon-induced correlations between electron phase factors, which is the
precursory mechanism for the detection of photon entanglement in distant
quantum interference devices. It is shown that the phase factors of the
electrons in interference experiments, which are initially independent of one
another, become correlated when the experiments are irradiated with correlated
photons. The setup considered here may also be useful in the general area of
flying and stationary qubits \cite{flying_qubits} and their interaction.

The rest of the paper is organized as follows. A possible implementation is
analyzed and background material is provided in section 2. The correlations
induced by the photons on the phase factors are quantified for the bipartite
case in section 3. The problem is approached through examples, that involve
classically and quantum mechanically correlated photons in number states and
in coherent states, in section 4. This is subsequently generalized to the
tripartite case in section 5, where examples are also provided. The results
are discussed and conclusions are drawn in section 6.

%----------------------------------------------------------------------------------------------------------
\section{Influence of entangled photons on distant interference experiments}
We begin by introducing the setup depicted in figure 1: two interference
devices for charged particles, \textbf{A} and \textbf{B}, are placed inside
cavities that are far from each other. A source $\rm S_{\rm EM}$ of two-mode
nonclassical microwaves sends one mode of frequency $\omega_1$ into the cavity
where \textbf{A} has been placed, and the other mode of frequency $\omega_2$
into the cavity where \textbf{B} has been placed. It has been shown that in
this case the correlation between the two electromagnetic field modes is
transferred to the distant quantum interference devices \cite{MY}. These
devices could be, for example, nanoscale superconducting quantum interference
devices (SQUIDs) \cite{Related_MY}, in which case the interfering particles
are Cooper pairs; or simply two-path electron interference devices \cite{MY}.
In either case the value of the phase factor, which depends on the external
electromagnetic fields, influences the measurable physical quantities (in the
case of superconducting rings the measurable variable is the current, while in
electron interference one measures the intensity of electrons on the
interference screen).

The external quantum fields are usually described by the vector potential
$\hat A_i$ and the electric field $\hat E_i$, which are dual quantum
variables. The $\hat A_i, \hat E_i$ can be transformed into another pair of
dual variables by integrating them around a small loop $l$ (that is, `small'
in comparison to the wavelength so that the field strengths are locally the
same). This operation yields the magnetic flux $\hat \phi =\oint _l \hat A_i
\rmd x_i$ and the electromotive force $\hat V_{\rm EMF}=\oint _l \hat E_i \rmd
x_i$, respectively. The boson creation and annihilation operators may now be
introduced as
\begin{eqnarray}
\hat a^{\dag}=\frac{1}{\sqrt{2}\xi}(\hat\phi-\rmi\omega^{-1}\hat V_{{\rm
EMF}}), \;\;\;\;\; \hat{a}= \frac{1}{\sqrt{2}\xi}\left(\hat{\phi} +
\rmi\omega^{-1} \hat{V}_{\rm EMF}\right) \label{creation_operator}
\end{eqnarray}
where $\xi$ is a constant proportional to the area enclosed by $l$. They obey
the usual commutation relation $[a,a^{\dagger}] = 1$ (note that we employ
units in which the Boltzmann constant, the Planck constant divided by $2\pi$,
and the speed of light in vacuum are set equal to one, $k_{\rm B}=\hbar=c=1$).
The flux operator is consequently written in the Heisenberg picture as
\begin{equation}
\hat {\phi}(t)= \exp (\rmi t{\cal H})\hat{\phi}(0)\exp (-\rmi t{\cal H})
\end{equation}
where
\begin{equation}
{\cal H}=H_{\rm free}+H_{\rm int},\;\;\;\;\; H_{\rm
free}=\omega(\hat{a}^{\dagger}\hat{a}+1/2).
\end{equation}
The full Hamiltonian ${\cal H}$ contains the free electromagnetic field
Hamiltonian and an interaction term $H_{\rm int}$, which includes the
Hamiltonian of the interfering charges as well. In this paper we assume that
the $H_{\rm int}$, which describes the back reaction from the charges to the
electromagnetic field, is neglected. In other words it is assumed that the
self-induced magnetic flux of the charges is negligible compared to the
external flux $\langle \hat {\phi}(t)\rangle$. In this approximation
\cite{Vourdas_AB,MY} we get
\begin{equation} \label{quantum_flux}
\hat \phi(t)=\frac{\xi}{\sqrt{2}} \left[\exp(\rmi \omega t)\hat{a}^\dagger +
\exp(-\rmi \omega t)\hat{a}\right].
\end{equation}
Exponentiating we obtain the phase factor for an electron of charge $\rme$:
\begin{equation} \label{q_definition}
\exp\left[\rmi \rme \hat \phi(t)\right]=D\left[\rmi q\exp(\rmi \omega
t)\right], \;\;\;\;\; q=\frac{\xi \rme}{\sqrt{2}}
\end{equation}
where $q$ is introduced as a scaled electric charge. $D(\lambda)\equiv
\exp(\lambda \hat{a}^{\dagger} - \lambda^{*}\hat{a})$ is the displacement
operator.

\begin{figure}
\begin{center}
\scalebox{0.5}{\includegraphics{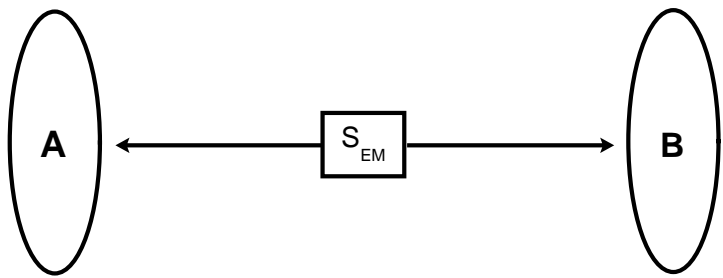}}
\end{center}
\caption{Two distant interference devices for charged particles, \textbf{A}
and \textbf{B}, are irradiated with nonclassical electromagnetic fields of
frequencies $\omega_1$ and $\omega_2$ correspondingly. The electromagnetic
fields emanate from a single source ${\rm S_{EM}}$ and are correlated. It is
required that the wavelengths of the fields are $\sim {\rm 1mm}$ (microwaves)
and that the interference devices have mesoscopic dimensions ($\sim 0.1\mu
{\rm m}$) operating at low temperatures of $10-100 {\rm mK}$, such that
$k_{\rm B}T \ll \hbar \omega_1,\hbar \omega_2$.}
\end{figure}

Let $\rho_{\rm A}$ be the density matrix describing the external nonclassical
electromagnetic field mode in cavity \textbf{A}. The expectation value of the
phase factor is given by the trace of the operator $\exp[\rmi \rme \hat
\phi_{\rm A}(t)]$ with respect to $\rho_{\rm A}$. It is easily seen that
taking the trace we obtain the single mode Weyl function
\begin{equation} \label{Weyl_A}
\tilde{W}_{\rm A}(\lambda_{\rm A}) \equiv \Tr [\rho_{\rm A} D(\lambda_{\rm
A})], \;\;\;\;\; \lambda_{\rm A}=\rmi q\exp(\rmi \omega_1 t).
\end{equation}
Similarly, the expectation value of the electron phase factor in experiment
\textbf{B} is given by the Weyl function
\begin{eqnarray}\label{Weyl_B}
\tilde{W}_{\rm B}(\lambda_{\rm B}) \equiv \Tr [\rho_{\rm B} D(\lambda_{\rm
B})], \;\;\;\;\; \lambda_{\rm B}=\rmi q\exp(\rmi \omega_2 t).
\end{eqnarray}
It is important to note that these `expectation values' are, in general,
complex numbers. The reason for this is that the operator $D(z)$ is not
Hermitian, since $D^{\dagger}(z)=D(-z)$.

To provide a physical interpretation consider that \textbf{A} is a two-path
electron interference experiment. With each path we associate a wavefunction
for the electrons, for example, $\psi_0$ and $\psi_1$ (let us assume equal
splitting among them, for simplicity). It has been shown elsewhere \cite{MY}
that the intensity, or number density, of electrons at position $x \equiv
\arg{\psi_0}-\arg{\psi_1}$ on the interference screen of experiment \textbf{A}
is given by
\begin{eqnarray}
\fl I_{\rm A}(x) = \Tr\left[\rho_{\rm A} |\psi_0 + \langle \exp(\rmi \rme
\hat{\phi}_{\rm A}) \rangle \psi_1| ^2 \right] =1 + |\tilde{W}_{\rm
A}(\lambda_{\rm A})|\cos\{x+\arg[\tilde{W}_{\rm A}(\lambda_{\rm A})]\}.
\end{eqnarray}
It is clearly seen that the absolute value of the expectation value of the
phase factor, $|\tilde W_{\rm A}(\lambda_{\rm A})|$, is the visibility
$\nu\equiv (I_{\rm max}-I_{\rm min})/(I_{\rm max}+I_{\rm min})$ of the
interference. The $\arg[\tilde W_{\rm A}(\lambda_{\rm A})]$ is the phase shift
induced on the electrons by the irradiating electromagnetic field.

%----------------------------------------------------------------------------------------------------------
\section{Correlations between electron phase factors}
In this section we show how the electron phase factors in distant interference
experiments become correlated when they are irradiated with correlated
photons. The nature of the correlation between the external photons can be
classical or quantum \cite{EN,EN_photons} and the aim here is to compare and
contrast the two cases. The difference between the two cases is firstly
clarified.

The photons of frequencies $\omega_1$ and $\omega_2$ are described by density
operators $\rho_{\rm A}$ and $\rho_{\rm B}$, correspondingly. If they are
completely independent of each other then the density operator describing the
bipartite state is factorizable, i.e., $\rho_{\rm fac}=\rho_{\rm A}\otimes
\rho_{\rm B}$. If they are classically correlated then the bipartite state is
described by the separable density operator $\rho_{\rm sep}=\sum_{k}P_{k}
\rho_{{\rm A},k}\otimes \rho_{{\rm B},k}$, where the $P_{k}$ are probabilities
that sum up to unity. If the two photons are quantum mechanically correlated
then their density operator $\rho_{\rm ent}$ is entangled and it can not be
cast in the above forms in any way.

The expectation values of the electron phase factors $\langle \exp(\rmi \rme
\hat\phi_{\rm A}) \rangle$ and $\langle \exp(\rmi \rme \hat\phi_{\rm B})
\rangle $ in the interference experiments \textbf{A} and \textbf{B} are given
by the single mode Weyl functions $\tilde {W}_{\rm A}(\lambda_{\rm A})$ and
$\tilde {W}_{\rm B}(\lambda_{\rm B})$ of equations (\ref{Weyl_A}) and
(\ref{Weyl_B}), correspondingly. It is also possible to measure the product of
the electron phase factors in \textbf{A} and \textbf{B} (joint phase factor).
The expectation value of this product, $\langle\exp(\rmi \rme \hat\phi_{\rm
A})\exp(\rmi \rme\hat\phi_{\rm B})\rangle$, is given by the two-mode Weyl
function
\begin{eqnarray} \label{Weyl_AB}
\tilde{W}_{\rm AB}(\lambda_{\rm A},\lambda_{\rm B}) =\Tr[\rho D(\lambda_{\rm
A})D(\lambda_{\rm B})].
\end{eqnarray}
In the case of independent subsystems, which are described by $\rho_{\rm
fac}=\rho_{\rm A}\otimes \rho_{\rm B}$, the $\tilde{W}_{\rm AB}(\lambda_{\rm
A},\lambda_{\rm B})$ is equal to the product $\tilde{W}_{\rm A}(\lambda_{\rm
A}) \tilde{W}_{\rm B}(\lambda_{\rm B})$. However for classically or quantum
mechanically correlated subsystems the two-mode Weyl function is not equal to
this product of one-mode Weyl functions, in general. This implies that the
electron phase factors in \textbf{A} and \textbf{B} are correlated with each
other.

In order to quantify the induced correlations between the electron phase
factors we define
\begin{eqnarray}\label{correlator}
C  \equiv  \tilde{W}_{\rm AB}(\lambda_{\rm A},\lambda_{\rm B}) -
\tilde{W}_{\rm A}(\lambda_{\rm A}) \tilde{W}_{\rm B}(\lambda_{\rm B}).
\end{eqnarray}
If the subsystems are not correlated with each other then $C = 0$. If they are
correlated then $C \neq 0$ (i.e., the real and imaginary parts of $C$ do not
both vanish).

The experimentally measurable quantities are the visibilities of the electron
interferences in \textbf{A} ($|\tilde{W}_{\rm A}(\lambda_{\rm A})|$) and in
\textbf{B} ($|\tilde{W}_{\rm B}(\lambda_{\rm B})|$); and the corresponding
shifts of the interference fringes $\arg(\tilde{W}_{\rm A}), \arg(
\tilde{W}_{\rm B})$. The absolute value (joint visibility) and the argument
(joint phase shift) of $\tilde{W}_{\rm AB}(\lambda_{\rm A},\lambda_{\rm B})$
have to be measured simultaneously in the two experiments. Alternatively one
may use a SQUID ring with a single Josephson junction irradiated with
nonclassical electromagnetic fields \cite{Related_MY}, in which case the
$\tilde{W}_{\rm A}$, $\tilde{W}_{\rm B}$ and the $\tilde{W}_{\rm AB}$ are
calculated from the expectation values of the currents in \textbf{A} and
\textbf{B} (and the product of the currents in both rings). For example, it is
known that the current measured in \textbf{A} is given by $I_{\rm A} = I_{\rm
cr} {\rm Im}[\tilde{W}_{\rm A}(\lambda_{\rm A})]$, where $I_{\rm cr}$ is the
critical current.

%----------------------------------------------------------------------------------------------------------
\section{Examples for the bipartite case}
In this section we consider particular examples of classically and quantum
mechanically correlated two-mode nonclassical electromagnetic fields in number
states and in coherent states. The fundamental relations that are necessary
for the derivation of the following results have been collected at the end of
the paper in appendix A (for the number states) and appendix B (for the
coherent states).

\subsection{Photons in number states}
Consider a two-mode electromagnetic field in the separable state
\begin{equation} \label{rho_sep_num}
\rho_{\rm sep} =\frac{1}{2}(|N_1 N_2 \rangle \langle N_1 N_2| + |N_2 N_1
\rangle \langle N_2 N_1 |).
\end{equation}
In this case the difference $C$ of equation (\ref{correlator}) is
\begin{equation} \label{C_sep_num}
\fl C_{\rm sep} = \exp(-q^2) L_{N_1}(q^2) L_{N_2}(q^2) - \frac{1}{4}
\exp(-q^2)  [L_{N_1}(q^2)+L_{N_2}(q^2)]^2
\end{equation}
where $L_{N}^{\alpha}$ are Laguerre functions. The $C_{\rm sep}$ is time
independent; it depends only on the number of photons $N_1,N_2$. It is clearly
seen that $|C|>0$ for any number of photons.

On the other hand the entangled number state $|n \rangle = 2^{-1/2} (|N_1 N_2
\rangle + |N_2 N_1 \rangle )$, with density operator
\begin{eqnarray}\label{rho_ent_num}
\rho_{\rm ent} = \rho_{\rm sep} + \frac{1}{2}(|N_1 N_2 \rangle \langle N_2
N_1| + |N_2 N_1 \rangle \langle N_1 N_2|)
\end{eqnarray}
yields
\begin{eqnarray}\label{C_ent_num}
C_{\rm ent} = C_{\rm sep} + \exp(-q^2) L_{N_1}^{N_2-N_1}(q^2)
L_{N_2}^{N_1-N_2}(q^2) \cos(\Omega t)
\end{eqnarray}
which is time dependent and oscillates around the $C_{\rm sep}$ with frequency
\begin{equation}\label{Omega}
\Omega=(N_1-N_2)(\omega_1-\omega_2).
\end{equation}
If there is no detuning between the external electromagnetic fields, in which
case $\omega_1=\omega_2$, then the $C_{\rm ent}$ is constant in time but it is
still different from the $C_{\rm sep}$. It is noted that for this example the
difference $C$ is purely real in both the separable and entangled cases.

\subsection{Photons in coherent states}
Consider two coherent states $|A_1\rangle$ and $|A_2\rangle$ in the
classically correlated state
\begin{eqnarray}\label{rho_sep_coherent}
\rho_{{\rm sep}}=\frac{1}{2}(|A_1 A_2\rangle \langle A_1 A_2| +|A_2 A_1\rangle
\langle A_2 A_1|).
\end{eqnarray}
In this case the reduced density operators that describe the coherent states
propagating in cavities \textbf{A} and \textbf{B} are
\begin{eqnarray}\label{reduced_rho_sep}
\rho_{\rm sep,A}=\rho_{\rm sep,B}=\frac{1}{2}(|A_1\rangle \langle A_1|
+|A_2\rangle\langle A_2|).
\end{eqnarray}

We also consider the entangled state $|S \rangle={\cal N}(|A_1 A_2\rangle
+|A_2 A_1\rangle)$ with density operator
\begin{eqnarray}\label{rho_ent_coherent}
\rho_{\rm ent}= 2{\cal N} ^2 \rho_{\rm sep}+ {\cal N}^2 (|A_1 A_2\rangle
\langle A_2 A_1| +|A_2 A_1\rangle \langle A_1 A_2|)
\end{eqnarray}
where the normalization constant, which is such that $\langle S |S\rangle =1$,
is given by
\begin{eqnarray}\label{normalization_constant}
{\cal N}=\left[2+2\exp\left(-|A_1-A_2|^2\right)\right]^{-1/2}.
\end{eqnarray}
In this case the reduced density operators in \textbf{A} and \textbf{B} are
\begin{eqnarray}\label{reduced_rho_ent}
\fl \rho_{\rm ent,A} = \rho_{\rm ent,B}= {\cal N}^2(|A_1\rangle \langle A_1|
+|A_2\rangle \langle A_2| + \tau_{12} |A_1\rangle\langle A_2| + \tau_{12}^{*}
|A_2\rangle\langle A_1|)
\end{eqnarray}
where
\begin{eqnarray} \label{tau_12}
\tau_{12} = \langle A_1|A_2\rangle = \exp\left(-\frac{|A_1|^2}{2}
-\frac{|A_2|^2}{2} + A_1^{*} A_2 \right).
\end{eqnarray}

The quantity $C$ of equation (\ref{correlator}) has been studied numerically
using the relations provided in appendix B. For the separable case we have
calculated numerically the $C_{\rm sep}=\tilde{W}_{\rm AB,sep}(\lambda_{\rm
A},\lambda_{\rm B}) - \tilde{W}_{\rm A,sep}(\lambda_{\rm A})\tilde{W}_{\rm
B,sep}(\lambda_{\rm B})$ and for the entangled case we have calculated the
$C_{\rm ent}=\tilde{W}_{\rm AB,ent}(\lambda_{\rm A},\lambda_{\rm B}) -
\tilde{W}_{\rm A,ent}(\lambda_{\rm A})\tilde{W}_{\rm B,ent}(\lambda_{\rm B})$.
These are complex quantities and therefore in the following we present the
results in terms of their absolute values $|C_{\rm sep}|,|C_{\rm ent}|$ and
their imaginary parts ${\rm Im}(C_{\rm sep}),{\rm Im}(C_{\rm ent})$.

%-------------
\begin{figure}
\begin{center}
\scalebox{0.5}{\includegraphics{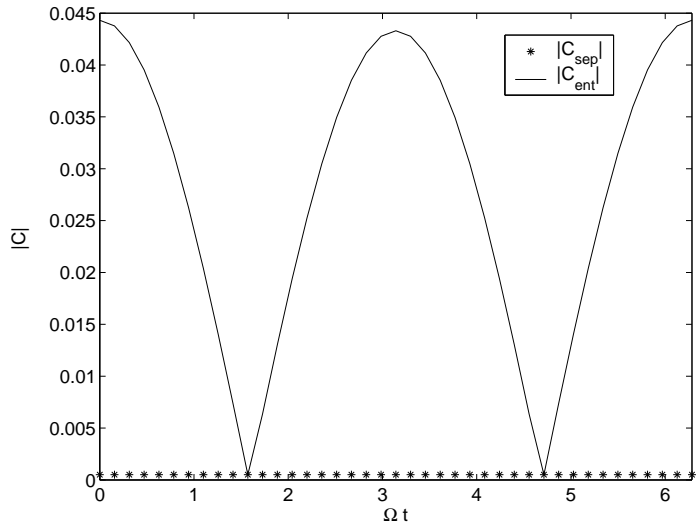}}
\end{center}
\caption{$|C_{\rm sep}|$ (line of stars) corresponding to the separable number
states of equation (\ref{rho_sep_num}) and $|C_{\rm ent}|$ (continuous line)
corresponding to the entangled number states of equation (\ref{rho_ent_num})
for $N_1=1,N_2=0$ as a function of $\Omega t$. The frequencies are
$\omega_1=1.2\times 10^{-4}$ and $\omega_2=10^{-4}$, in units where $k_{\rm
B}=\hbar=c=1$. Note that $|C_{\rm sep}|$ is not zero but $5\times 10^{-4}$.}
\end{figure}

\subsection{Numerical results}
For the numerical results in this section the values of the microwave
frequencies have been set at $\omega_1=1.2\times 10^{-4}$ and
$\omega_2=1.0\times 10^{-4}$. We have used units in which $k_{\rm
B}=\hbar=c=1$. Other fixed parameters are $\xi=1$ and the dimensionless
electric charge $\rme=(4\pi/137)^{1/2}$.

We study the entangled number state $|\beta \rangle = 2^{-1/2} (|01 \rangle +
|10 \rangle)$ and its closest separable state. We therefore let $N_1=1,N_2=0$
in the separable state of equation (\ref{rho_sep_num}) and the entangled state
of equation (\ref{rho_ent_num}). The corresponding results for the $|C_{\rm
sep}|$ and the $|C_{\rm ent}|$ against $\Omega t$ have been plotted in figure
2. We note that the $|C_{\rm sep}|$, which is time independent, is not zero
but it is very small in this case ($\simeq 5\times 10^{-4}$).

In the case of coherent states we study the separable state of equation
(\ref{rho_sep_coherent}) and the entangled state of equation
(\ref{rho_ent_coherent}) for the same average number of photons as in the
number states, that is, $|A_1|^2=N_1$ and $|A_2|^2=N_2$ (whereas
$\arg{A_1}=0,\arg{A_2}=0$). The results for the $|C_{\rm sep}|$ and the
$|C_{\rm ent}|$ have been plotted against $\Omega t$ in subplots (a) and (c)
of figure 3, correspondingly. We note that in this case $C$ is complex and
also $C_{\rm sep}$ is time dependent (in contrast to the case of number
states). In subplots (b) and (d) of figure 3 the imaginary parts, ${\rm
Im}(C_{\rm sep})$ and ${\rm Im}(C_{\rm ent})$, have been plotted against
$\Omega t$.

In figure 2 we see that both the $C_{\rm sep}$ and the $C_{\rm ent}$ are
nonzero; and that the $C_{\rm ent}$ is time dependent. In fact this is true
for any number of photons $N_1,N_2$ in the separable and entangled states
$\rho_{\rm sep},\rho_{\rm ent}$ as we can see from equations (\ref{C_sep_num})
and (\ref{C_ent_num}). Consequently the electron phase factors become
correlated when the interference devices are irradiated with classically
correlated ($\rho_{\rm sep}$) or quantum mechanically correlated ($\rho_{\rm
ent}$) photons in number states. Clearly the quantity $C$ of equation
(\ref{correlator}) is different for the two cases, which implies that the
nature of the correlation between the irradiating photons influences the
induced correlation between the topological phase factors. In figure 3 we see
that the same general result is true for the case of classically and quantum
mechanically correlated photons in coherent states. It is also evident that
the correlations between the phase factors are influenced by the quantum noise
and statistics of the external photons, by comparison of figures 2 and 3,
which correspond to photons in number states and coherent states.

%-------------
\begin{figure}
\begin{center}
\scalebox{0.5}{\includegraphics{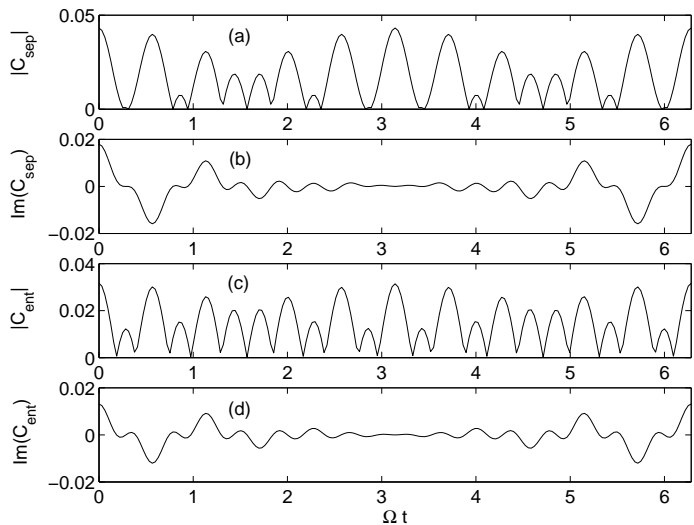}}
\end{center}
\caption{(a) $|C_{\rm sep}|$ and (b) ${\rm Im}(C_{\rm sep})$ corresponding to
the separable coherent states of equation (\ref{rho_sep_coherent}); (c)
$|C_{\rm ent}|$ and (d) ${\rm Im}(C_{\rm ent})$ corresponding to the entangled
coherent states of equation (\ref{rho_ent_coherent}) for $A_1=1,A_2=0$ as a
function of $\Omega t$. The frequencies are $\omega_1=1.2\times 10^{-4}$ and
$\omega_2=10^{-4}$, in units where $k_{\rm B}=\hbar=c=1$.}
\end{figure}

%----------------------------------------------------------------------------------------------------------
\section{Examples for the tripartite case}
In this section we consider three electron interference devices of mesoscopic
dimensions that are placed inside distant microwave cavities. The interference
experiments \textbf{A}, \textbf{B}, and \textbf{C} are irradiated with
nonclassical electromagnetic fields of frequencies $\omega_1, \omega_2$, and
$\omega_3$, correspondingly. The three electromagnetic field modes are
described by density operators $\rho_{\rm A}, \rho_{\rm B}$, and $\rho_{\rm
C}$. If they are completely independent of each other then the density
operator describing the tripartite state is factorizable, i.e., $\rho_{\rm
fac}=\rho_{\rm A}\otimes \rho_{\rm B} \otimes \rho_{\rm C}$. If they are
classically correlated then the tripartite state is described by the separable
density operator $\rho_{\rm sep}=\sum_{k}P_{k} \rho_{{\rm A},k}\otimes
\rho_{{\rm B},k} \otimes \rho_{{\rm C},k}$. If the three field modes are
quantum mechanically correlated then their density operator $\rho_{\rm ent}$
is entangled and it can not be written in a separable form.

The phase factor acquired by the interfering electrons in \textbf{A} is given
by $\tilde{W}_{\rm A}(\lambda_{\rm A})$ of equation (\ref{Weyl_A}) and the
phase factor in \textbf{B} is given by $\tilde{W}_{\rm B}(\lambda_{\rm B})$ of
equation (\ref{Weyl_B}). Similarly the phase factor in \textbf{C} is obtained
from $\tilde{W}_{\rm C}(\lambda_{\rm C})=\Tr[\rho_{\rm C}D(\lambda_{\rm C})]$,
where $\lambda_{\rm C}=\rmi q\exp(\rmi \omega_3 t)$. We can also measure the
product of the three phase factors, which is given by the three mode Weyl
function
\begin{eqnarray} \label{Weyl_ABC}
\tilde{W}_{\rm ABC}(\lambda_{\rm A},\lambda_{\rm B},\lambda_{\rm C}) =
\Tr[\rho D(\lambda_{\rm A})D(\lambda_{\rm B})D(\lambda_{\rm C})].
\end{eqnarray}
The tripartite correlations between the electron phase factors can be
quantified with a straightforward generalization of the quantity $C$ of
equation (\ref{correlator}), that is, in this case we define
\begin{eqnarray}\label{c3}
C  \equiv  \tilde{W}_{\rm ABC}(\lambda_{\rm A},\lambda_{\rm B}, \lambda_{\rm
C}) - \tilde{W}_{\rm A}(\lambda_{\rm A}) \tilde{W}_{\rm B}(\lambda_{\rm B})
\tilde{W}_{\rm C}(\lambda_{\rm C}).
\end{eqnarray}
If the phase factors are not correlated then $C = 0$. If they are correlated
then $|C|> 0$ (and also possibly ${\rm Im}(C)\neq 0$).

\subsection{Photons in number states}
As an example of tripartite number states consider the separable state
\begin{equation} \label{rho3_sep_num}
\rho_{\rm sep} =\frac{1}{2}(|N_1 N_2 N_3 \rangle \langle N_1 N_2 N_3| + |N_2
N_3 N_1 \rangle \langle N_2 N_3 N_1 |)
\end{equation}
and the entangled state $|n_{\rm tri} \rangle = 2^{-1/2} (|N_1 N_2 N_3 \rangle
+ |N_2 N_3 N_1 \rangle )$ with density operator
\begin{eqnarray}\label{rho3_ent_num}
\rho_{\rm ent} = \rho_{\rm sep} + \frac{1}{2}(|N_1 N_2 N_3\rangle \langle N_2
N_3 N_1| + |N_2 N_3 N_1 \rangle \langle N_1 N_2 N_3|).
\end{eqnarray}

The results for the three mode Weyl function of equation (\ref{Weyl_ABC})
corresponding to the separable and entangled number states are
straightforward, albeit lengthy. Only the numerical calculations are presented
in terms of time $\Omega ' t$, where $\Omega '$ has replaced $\Omega$ of
equation (\ref{Omega}), which was valid for the bipartite case. In particular
it is not hard to show that the difference between the separable and entangled
Weyl functions includes a time dependent term of frequency $\Omega '$, which
is given by
\begin{eqnarray}
\fl \tilde{W}_{\rm ABC,ent}-\tilde{W}_{\rm ABC,sep} \propto {\rm Re}(\langle
N_1|D(\lambda_{\rm A})|N_2 \rangle \langle N_2|D(\lambda_{\rm B})|N_3 \rangle
\langle N_3|D(\lambda_{\rm C})|N_1 \rangle ).
\end{eqnarray}
From this term we obtain the appropriate frequency for the tripartite case,
namely
\begin{equation}
\Omega ' = N_1 (\omega_3-\omega_1) + N_2 (\omega_1-\omega_2) +
N_3(\omega_2-\omega_3).
\end{equation}

%-------------
\begin{figure}
\begin{center}
\scalebox{0.5}{\includegraphics{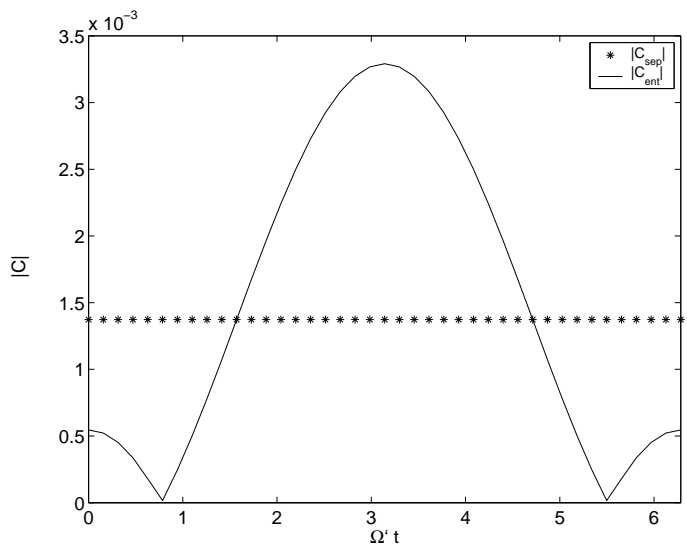}}
\end{center}
\caption{$|C_{\rm sep}|$ (line of stars) corresponding to the separable number
states of equation (\ref{rho3_sep_num}) and $|C_{\rm ent}|$ (continuous line)
corresponding to the entangled number states of equation (\ref{rho3_ent_num})
for $N_1=0,N_2=1,N_3=2$ as a function of $\Omega ' t$. The frequencies are
$\omega_1=1.2\times 10^{-4}$, $\omega_2 = 1.1\times 10^{-4}$, and $\omega_3 =
1.0\times 10^{-4}$ in units where $k_{\rm B}=\hbar=c=1$.}
\end{figure}

%-------------
\begin{figure}
\begin{center}
\scalebox{0.5}{\includegraphics{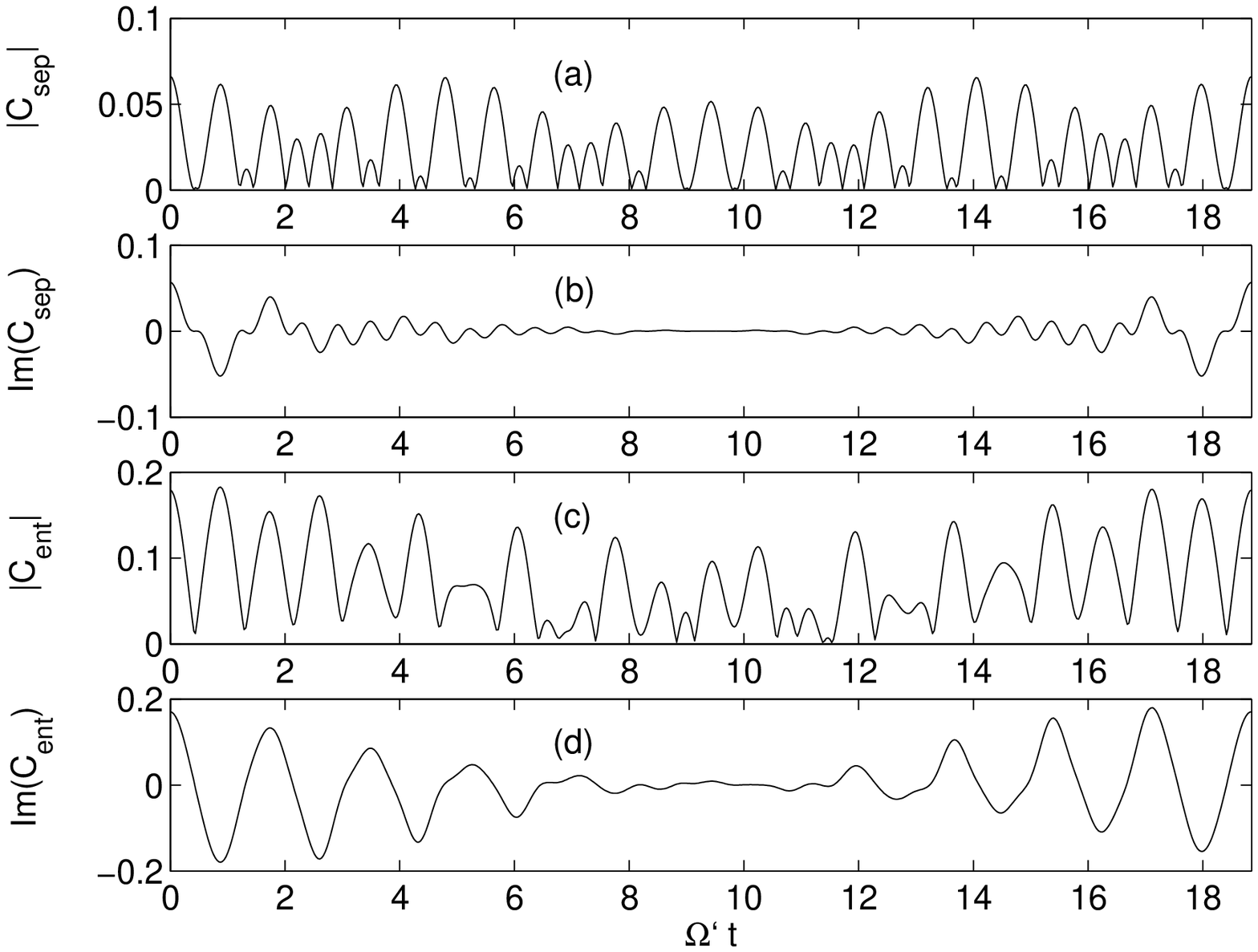}}
\end{center}
\caption{(a) $|C_{\rm sep}|$ and (b) ${\rm Im}(C_{\rm sep})$ corresponding to
the separable coherent states of equation (\ref{rho3_sep_coherent}); (c)
$|C_{\rm ent}|$ and (d) ${\rm Im}(C_{\rm ent})$ corresponding to the entangled
coherent states of equation (\ref{rho3_ent_coherent}) for
$A_1=0,A_2=1,A_3=\sqrt{2}$ as a function of $\Omega' t$. The frequencies are
$\omega_1=1.2\times 10^{-4}$, $\omega_2 = 1.1\times 10^{-4}$, and $\omega_3 =
1.0\times 10^{-4}$ in units where $k_{\rm B}=\hbar=c=1$.}
\end{figure}

\subsection{Photons in coherent states}
We consider the separable coherent state
\begin{eqnarray}\label{rho3_sep_coherent}
\rho_{{\rm sep}}=\frac{1}{2}(|A_1 A_2 A_3 \rangle \langle A_1 A_2 A_3| +|A_2
A_3 A_1\rangle \langle A_2 A_3 A_1|).
\end{eqnarray}
In this case the reduced density operators are
\begin{eqnarray}\label{3reduced_rho_sep}
\rho_{\rm sep,A} &=& 2^{-1}(|A_1\rangle \langle A_1| +|A_2\rangle\langle
A_2|) \\
\rho_{\rm sep,B} &=& 2^{-1}(|A_2\rangle \langle A_2| +|A_3\rangle\langle
A_3|) \nonumber \\
\rho_{\rm sep,C} &=& 2^{-1}(|A_3\rangle \langle A_3| +|A_1\rangle\langle
A_1|). \nonumber
\end{eqnarray}

We also consider the entangled state $|S_{\rm tri} \rangle={\cal N '}(|A_1 A_2
A_3 \rangle +|A_2 A_3 A_1 \rangle)$ with density operator
\begin{eqnarray}\label{rho3_ent_coherent}
\rho_{\rm ent}= 2{\cal N'} ^2 \rho_{\rm sep}+ {\cal N'}^2 (|A_1 A_2 A_3\rangle
\langle A_2 A_3 A_1| +|A_2 A_3 A_1\rangle \langle A_1 A_2 A_3|)
\end{eqnarray}
where the normalization constant is given by
\begin{eqnarray}
{\cal N '}=[2+2{\rm Re}(\tau_{12} + \tau_{23} + \tau_{31}) ]^{-1/2}
\end{eqnarray}
for $\tau_{ij} = \langle A_i|A_j\rangle = \exp(-|A_i|^2/2 -|A_j|^2/2 + A_i^{*}
A_j)$ as in equation (\ref{tau_12}), for example. In this case the reduced
density operators are
\begin{eqnarray}\label{3reduced_rho_ent}
\rho_{\rm ent,A} &=& {\cal N'}^2(2\rho_{\rm sep,A} + \tau_{13}\tau_{32}
|A_1\rangle\langle A_2| + \tau_{13}^{*}\tau_{32}^{*} |A_2\rangle\langle A_1|)
\\
\rho_{\rm ent,B} &=& {\cal N'}^2(2\rho_{\rm sep,B} + \tau_{21}\tau_{13}
|A_2\rangle\langle A_3| + \tau_{21}^{*}\tau_{13}^{*} |A_3\rangle\langle A_2|)
\nonumber \\
\rho_{\rm ent,C} &=& {\cal N'}^2(2\rho_{\rm sep,C} + \tau_{12}\tau_{23}
|A_3\rangle\langle A_1| + \tau_{12}^{*}\tau_{23}^{*} |A_1\rangle\langle A_2|).
\nonumber
\end{eqnarray}

%-------------
\begin{figure}
\begin{center}
\scalebox{0.5}{\includegraphics{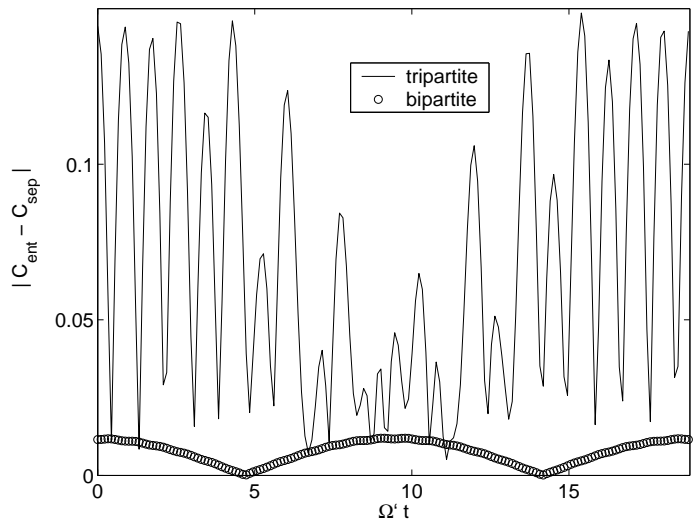}}
\end{center}
\caption{$|C_{\rm ent} - C_{\rm sep}|$ for coherent states with
$A_1=0,A_2=1,A_3=\sqrt{2}$ as a function of $\Omega' t$. The solid line
corresponds to the tripartite case of the separable and entangled states of
equations (\ref{rho3_sep_coherent}), (\ref{rho3_ent_coherent}). The line of
circles corresponds to the bipartite case of the separable and entangled
states of equations (\ref{rho_sep_coherent}), (\ref{rho_ent_coherent}). The
frequencies are $\omega_1=1.2\times 10^{-4}$, $\omega_2 = 1.1\times 10^{-4}$,
and $\omega_3 = 1.0\times 10^{-4}$ in units where $k_{\rm B}=\hbar=c=1$.}
\end{figure}

\subsection{Numerical results}
For the numerical results in this section the photon frequencies are
$\omega_1=1.2\times 10^{-4}$, $\omega_2 = 1.1\times 10^{-4}$, $\omega_3 =
1.0\times 10^{-4}$ in units where $k_{\rm B}=\hbar=c=1$, and $\xi=1$.

We study the entangled tripartite state $|\beta_{\rm tri} \rangle = 2^{-1/2}
(|012 \rangle + |120 \rangle)$ and its closest separable state. We therefore
let $N_1=0,N_2=1$, and $N_3=2$ in the separable number state of equation
(\ref{rho3_sep_num}) and the entangled number state of equation
(\ref{rho3_ent_num}). The corresponding results for the $|C_{\rm sep}|$ and
the $|C_{\rm ent}|$ against $\Omega ' t$ in the case of tripartite number
states have been plotted in figure 4. In this case both the $|C_{\rm sep}|$
and the $|C_{\rm ent}|$ are very small, but in principle measurable.

In the case of coherent states we study the separable state of equation
(\ref{rho3_sep_coherent}) and the entangled state of equation
(\ref{rho3_ent_coherent}) for the same average number of photons as in the
number states, therefore we let $A_1=0,A_2=1,A_3=\sqrt{2}$. The $|C_{\rm
sep}|$ and the $|C_{\rm ent}|$ have been plotted against $\Omega ' t$ in (a)
and (c) of figure 5, correspondingly. In subplots (b) and (d) of figure 5 the
corresponding imaginary parts, ${\rm Im}(C_{\rm sep})$ and ${\rm Im}(C_{\rm
ent})$, have been plotted against $\Omega ' t$.

In figure 6 we show the $|C_{\rm ent} - C_{\rm sep}|$ for coherent states with
$A_1=0,A_2=1,A_3=\sqrt{2}$ as a function of $\Omega' t$. The solid line
corresponds to the tripartite case, where the $\rho_{\rm sep}$ and $\rho_{\rm
ent}$ are given by equations (\ref{rho3_sep_coherent}) and
(\ref{rho3_ent_coherent}), respectively. The line of circles corresponds to
the bipartite case, where the $\rho_{\rm sep}$ and $\rho_{\rm ent}$ are given
by equations (\ref{rho_sep_coherent}) and (\ref{rho_ent_coherent}),
respectively. It is clearly seen that in both cases there is a significant
difference between the $C_{\rm sep}$ and the $C_{\rm ent}$. It is also seen
that the absolute value of $C_{\rm ent} - C_{\rm sep}$ for the tripartite case
is an order of magnitude greater than in the bipartite case. Therefore the
quantum part of $C$ does not diminish as the photon correlations are
distributed to more than two electron interference devices.

%----------------------------------------------------------------------------------------------------------
\section{Discussion}
It has been recognized that geometrical and topological phases
\cite{Geometric_phases,AB,AC,Wilkens_phase} could be harnessed for the
purposes of inherently fault-tolerant quantum computation \cite{AB_gates,
Kitaev}. It has also been known for some time that the quantum mechanical
correlations of physical states are a useful resource for quantum information
processing \cite{EN}. The aim of this paper has been to study the
photon-induced correlations of topological phase factors for charged particles
in distant interference experiments. It has been shown that the classical or
quantum correlations of the irradiating photons are transferred to the phase
factors of the circulating electrons. This mechanism may allow for the
detection of photon entanglement using nanoscale electronic devices
\cite{Related_MY,MY}.

In particular, we have considered the one-mode Weyl functions of equations
(\ref{Weyl_A}) and (\ref{Weyl_B}) for the density operators $\rho_{\rm A}$ and
$\rho_{\rm B}$ of the photons propagating in the distant cavities \textbf{A}
and \textbf{B}. They yield the expectation values of the electron phase
factors in the two interference experiments. These can be measured
experimentally through the visibility and the phase shift of the interference
fringes. We have also considered the two-mode Weyl function of equation
(\ref{Weyl_AB}) for the bipartite state $\rho$. This yields the joint phase
factor in both experiments. Using these Weyl functions we have defined the
difference $C$ of equation (\ref{correlator}), which vanishes only for
independent subsystems. Considering suitable examples of classically and
quantum mechanically correlated photons in number and in coherent states, we
have shown that $C$ does not vanish and that therefore the electron phase
factors are correlated. We have also shown that the value of $C$ depends on
the quantum noise and statistics of the external photons (figures 2 and 3).
Further work is required in order to distinguish between classical and quantum
mechanical correlations using the proposed setup. One possibility would be to
derive a Bell-type inequality for the two-mode Weyl function, which is obeyed
in the separable case, but it is violated in the entangled case.

It has also been shown that the same general result applies to the tripartite
case. In this case the joint phase factor is measured in three distant
electron interference experiments and its expectation value is given by the
three mode Weyl function of equation (\ref{Weyl_ABC}). The difference $C$ is
in this case replaced by that of equation (\ref{c3}). Numerical results have
been presented in figures 4-6 for several examples of classically and quantum
mechanically correlated number states and coherent states.

In conclusion we have shown that it is possible to entangle the topological
phase factors of interfering electrons that are irradiated with nonclassical
electromagnetic fields. In future work it would be very interesting to derive
similar results on the photon-induced entanglement of geometric phases
acquired by spin-$1/2$ particles \cite{Guridi}, or Cooper pairs in mesoscopic
Josephson junctions \cite{Shao}, for example. In the last few years there has
been a lot of work on the role of entanglement in mesoscopic devices
\cite{mesoscopic_ent}. The setup discussed in this paper may be useful in the
production of entangled electric charges in a normal conductor or a
superconductor using topological phases that are induced by external photons.
This is within the realm of current experimental techniques, whereby a
nanoscale Josephson device can be controlled with a single microwave photon
\cite{experiment}.

%----------------------------------------------------------------------------------------------------------
\appendix
\section{Relations for number states}
The following relation yields the matrix elements of the displacement operator
in the number state basis \cite{Roy}:
\begin{eqnarray} \label{matrix_elements_D}
\langle m |D(z)|n \rangle = \left(\frac{n!}{m!}\right)^{1/2} z^{m-n} \exp
\left(\frac{-|z|^2}{2}\right) L_{n}^{m-n}(|z|^2).
\end{eqnarray}
Using this it can easily be shown that
\begin{eqnarray}
\tilde{W}_{\rm A}(\lambda_{\rm A})=\tilde{W}_{\rm B}(\lambda_{\rm B}) =
2^{-1}\exp(-q^2/2)[L_{N_1}(q^2)+L_{N_2}(q^2)]
\end{eqnarray}
for the $\rho_{\rm sep}$ of equation (\ref{rho_sep_num}) and the $\rho_{\rm
ent}$ of equation (\ref{rho_ent_num}). The two-mode Weyl function of equation
(\ref{Weyl_AB}) for the $\rho_{\rm sep}$ is
\begin{eqnarray}
\tilde{W}_{\rm AB,sep}(\lambda_{\rm A},\lambda_{\rm B}) = \exp(-q^2)
L_{N_1}(q^2)L_{N_2}(q^2).
\end{eqnarray}
However for the $\rho_{\rm ent}$ we have
\begin{eqnarray}
\fl \tilde{W}_{\rm AB,ent}(\lambda_{\rm A},\lambda_{\rm B}) =  \tilde{W}_{\rm
AB,sep}(\lambda_{\rm A},\lambda_{\rm B}) + \exp(-q^2) L_{N_1}^{N_2-N_1}(q^2)
L_{N_2}^{N_1-N_2}(q^2) \cos(\Omega t).
\end{eqnarray}

\section{Relations for coherent states}
In the coherent states basis we have
\begin{eqnarray}
\langle A |D(z)|B\rangle = \langle 0|D(-A+z+B)|0\rangle\exp(\chi)
\end{eqnarray}
where the $\langle 0|D(-A+z+B)|0\rangle$ can be calculated with the help of
equation (\ref{matrix_elements_D}) and the phase $\chi$ is given by
\begin{eqnarray}
\chi = \frac{1}{2}(-Az^* +A^*z - AB^* +A^*B  -z^* B +z B^*)
\end{eqnarray}
for any complex numbers $A,B,z$.

\ack{The author would like to thank Prof A. Vourdas (University of Bradford)
for critical comments and guidance, Dr J. K. Pachos (University of Cambridge)
for stimulating discussions, and Prof M. Babiker (University of York) for
support and encouragement.}

%----------------------------------------------------------------------------------------------------------
%-------------------------------------- REFERENCES --------------------------------------------------------

%----------------------------------------------------------------------------------------------------------
\end{document}